\documentstyle{mn}

\def\rchi{{${\chi}_{\nu}^{2}$}}

\title{Spectroscopic observations of the eclipsing Polar MN Hya (RX 
J0929--24)}
\author[G. Ramsay \& P. J. Wheatley]
{Gavin Ramsay$^{1}$ \& Peter J Wheatley$^{2}$\\
$^{1}$Mullard Space Science Laboratory, University College London,
Holmbury St.Mary, Dorking, Surrey, RH5 6NT\\
$^{2}$X-ray Astronomy Group, Dept of Physics \& Astronomy, University
of Leicester, University Road, Leicester, LE1 7RH}

\date{Accepted {\sl MNRAS} July 1998}

\begin{document}

\maketitle

\begin{abstract} 

We present low--medium resolution optical spectroscopy of the
eclipsing AM Her system MN Hya (RX J0929--24). We determine the
magnetic field strength at the primary accretion region of the white
dwarf to be 42MG from the spacing of cyclotron features visible during
$\phi\sim$0.4--0.7. From spectra taken during the eclipse we find that
the secondary has a M3--4 spectral type. Combined with the eclipse
photometry of Sekiguchi, Nakada \& Bassett and an estimate of the
interstellar extinction we find a distance of $\sim$300--700pc. We
find unusual line variations at $\phi\sim$0.9: H$\alpha$ is seen in
absorption and emission. This is at the same point in the orbital
phase that a prominent absorption dip is seen in soft X-rays.

\end{abstract}


\vspace{2.5cm}

\section{Introduction}

MN Hya (RX J0929.1--2404) was discovered during the {\sl ROSAT}
all-sky survey and subsequently found to be an Polar (or AM Her) type
of Cataclysmic Variable (Sekiguchi, Nakada \& Bassett 1994 and Buckley
et al 1998a). Polars are interacting binary stars in which the white
dwarf primary has a magnetic field strength sufficiently high to
synchronise its spin period with the binary orbital period. Further,
the magnetic field prevents the formation of an accretion disc and the
accretion stream impacts directly onto the surface of the white dwarf.

MN Hya is one of 9 currently known eclipsing AM Her systems. These are
particularly important systems to study since parameters such as the
masses of the component stars can, in principle, be accurately
determined. Of these systems, three (including MN Hya at
$P_{orb}$=3.39 hrs) are above the upper edge of the `period gap'
($\sim$2--3 hrs). MN Hya also shows two `dips' in soft X-rays where the
accretion stream obscures emission from the hot post shock region
above the white dwarf (Buckley et al 1998b). The deepest dip occurs at
$\phi\sim$0.9 (just before the eclipse of the white dwarf by the
secondary star), while another less deep dip occurs at
$\phi\sim$0.5. These dips are not seen in the optical continuum
(Sekiguchi, Nakada \& Bassett 1994).

We have obtained low-medium resolution phase resolved optical
spectroscopy of MN Hya and determined the spectral type of the
secondary star, its distance, the magnetic field strength of white
dwarf and examine how the optical spectrum varies through the
absorption dips.

\section{Observations and Reduction}

Spectra were obtained using the ESO 3.6m telescope at La Silla, Chile
on 1997 Feb 7/8 \& 8/9. The conditions were photometric and the seeing
was between 1.0--1.5$^{''}$. On the first night alternate images (with
exposures typically 300 sec) were taken with the R300 (6000--9800 \AA)
and B300 (3800--6800 \AA) grisms. With a slit of 1.5$^{''}$ the
resolution in both grisms was $\sim$15\AA. On the second night, only
the B300 grism was used and the exposures were 180 sec. Wavelength
calibration of the flat-fielded and bias-subtracted two-dimensional
images was performed using a He-Ar arc spectra taken at the start and
end of the night (flexure is negligible on ESO 3.6m telescope). The
wavelength calibrated spectra were flux calibrated using the standard
stars GD 108 and Feige 67.

\section{The strength of the magnetic field of the white dwarf}

One method of determining the magnetic field strength of the white
dwarf is from the spacing of cyclotron harmonics. For the field
strengths seen in Polars ($B\sim$10--200 MG) cyclotron harmonics are
seen as broad humps in optical/IR spectra. The cyclotron flux
originates from electrons spiraling around the field lines in the
post-shock region above the surface of the white dwarf. These humps
vary in their intensity and (to a lesser degree their wavelength) as a
function of our viewing angle to the post-shock region (ie they vary
over the orbital phase).

\begin{figure*}
\begin{center}
\setlength{\unitlength}{1cm}
\begin{picture}(8,22)
\put(-6,-2.){\includegraphics{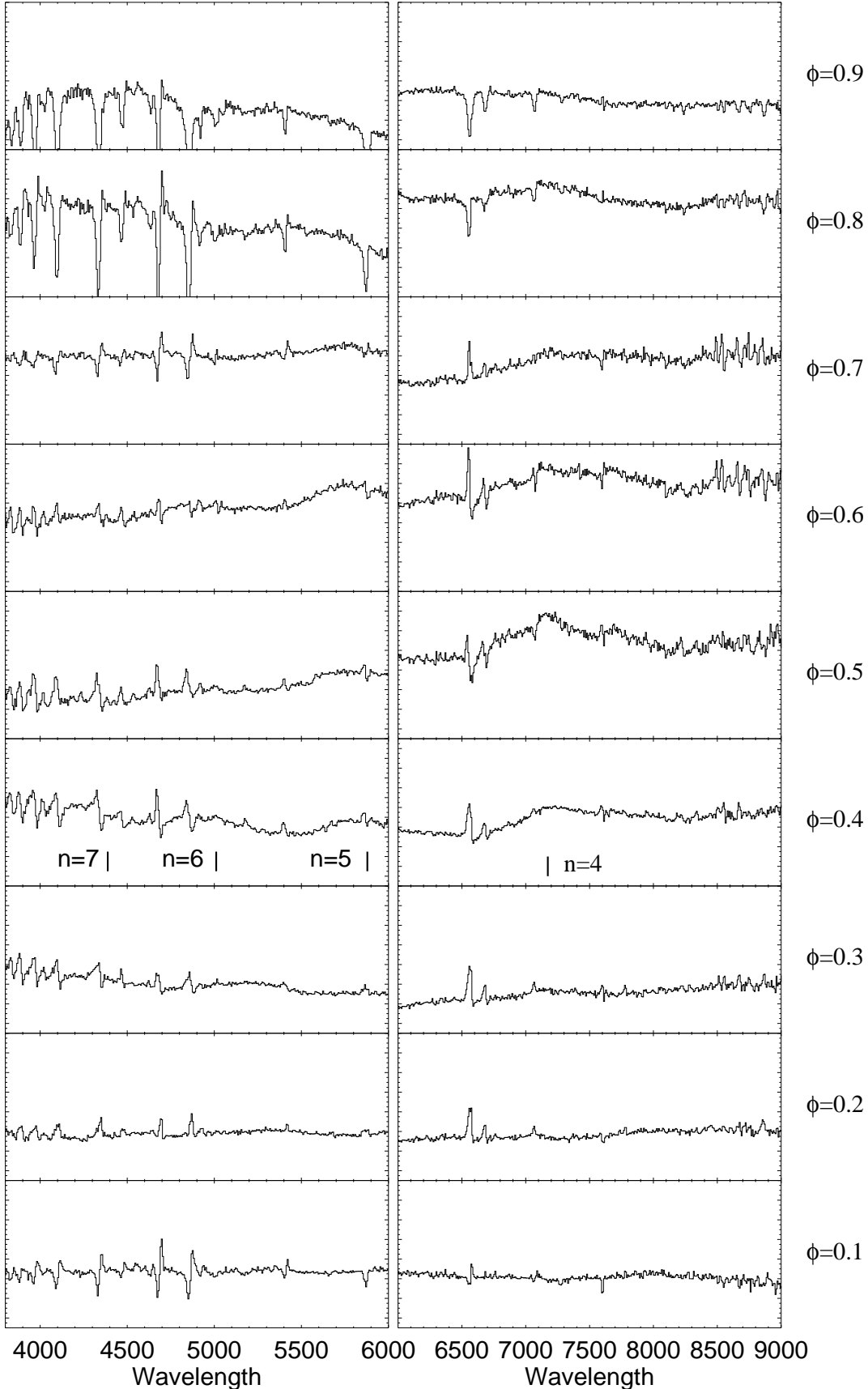}}
\end{picture}
\end{center}
\caption{The phase averaged normalised spectra covering the binary orbital
cycle (the eclipse phase is shown to the right).  Averaged blue and
red spectra covering 0.1 of the orbital cycle were obtained and then
divided by the mean blue and red spectrum respectively. (For example
the bin $\phi$=0.1 covers $\phi$=0.05--0.15). The unusual line
profiles are simply due to radial velocity variations over the orbital
cycle. The vertical axis covers the same interval range.  The
wavelengths of harmonic numbers $n=4-7$ are shown for a temperature of
9keV, a magnetic field strength of 42MG and viewing angle of
90$^{\circ}$.}
\label{cychumps} 
\end{figure*}

There was no obvious sign of cyclotron features in our optical
spectra. To make a more detailed search for such features we made a
mean `blue' and `red' spectrum: the mean blue spectrum was made up of
`blue' spectra (3800--6800 \AA) taken on the second night, while the
mean red spectrum was made up of `red' spectra (6000--9800 \AA) taken
on the first night (on the second night no red spectra were
obtained). We then averaged spectra to give spectra covering the phase
range $\phi$=0.05--0.15, 0.15--0.25 etc and then normalised these
spectra by dividing by the appropriate mean spectrum (we use the
ephemeris of Buckley et al 1998a which defines $\phi$=0.0 as the
center of the eclipse of the white dwarf by the secondary). Figure
\ref{cychumps} shows these normalised spectra over the binary orbit.
There is no evidence for cyclotron features until the phase interval
covering $\phi$=0.35--0.45 when humps are seen at $\sim$ 5800 \&
7200\AA. These continue to be seen until $\phi\sim$0.7. This is
confirmed if we divide the spectrum covering $\phi$=0.35--0.45 by the
spectrum covering $\phi$=0.05--0.15.

To determine the wavelength of cyclotron harmonics for a given
magnetic field strength $B$, we write $\omega_{B}=eB/m_{e}c$ (where
$\omega_{B}$ is the fundamental cyclotron frequency) as
$\omega_{B}=1.76\times10^{14}B_{7}$ (where $B_{7}=B/10^{7}$G).  We
then use equation (3) of Cropper et al (1988) and put it in terms of
$\lambda$:

\begin{equation}
\frac{\omega}{\omega_{B}}=\frac{-1+\sqrt{1+(8n\sin^{2}\theta/\mu)}}
{(4\sin^{2}\theta/\mu)}=\frac{\lambda_{B}}{\lambda}
\end{equation}

\noindent where $\mu=m_{e}c^{2}/kT=511.1/T$, $T$ is in units of keV,
$n$ is the cyclotron harmonic number and $\theta$ is the viewing angle
to the magnetic field angle.  We can now rearrange the above as a
function of $\lambda$:

\begin{equation}
\lambda = \frac{4.28\times10^{5}\sin^{2}\theta} {\mu B_{7}
\bigl(-1+\sqrt{1+(8n\sin^{2}\theta/\mu)}\bigr) }
\end{equation}

\noindent where $\lambda$ is in \AA.

To determine the magnetic field strength we searched a range of
parameters which gave consecutive harmonics placed at $\sim$ 5800 \&
7200\AA. For lower values of $\theta$ the harmonics are blue shifted
by a small amount compared to higher values. For low magnetic fields
($B\sim$15MG) the individual harmonics become indistinguishable and
form a continuum. Similarly for temperatures $>$15keV individual
harmonics are not visible. By comparing the expected wavelength of the
harmonics by eye to the observed spectrum, we found that $B$=42MG,
$kT$=9keV and $\theta$=90$^{\circ}$ gave the best `fit'. Assuming that
we have correctly identified harmonics at $\sim$ 5800 \& 7200\AA, we
estimate the error on $B$ to be $\sim$1--2MG based on our search of
the variable parameters.

The value of $B$=42MG found here compares with $B\sim$20MG estimated
by Buckley et al (1998a) from a single spectrum of exposure 3000
sec. Their spectrum was not flux-calibrated and they caution that the
small-scale variations seen in their spectrum could have been due to
other factors rather than cyclotron humps. We therefore consider that
our estimate of $B$=42MG to be more reliable. We will examine this in
more detail in \S \ref{daveb}.

\section{The secondary star and its distance}

\begin{figure}
\begin{center}
\setlength{\unitlength}{1cm}
\begin{picture}(8,13)
\put(-1,-1){\includegraphics{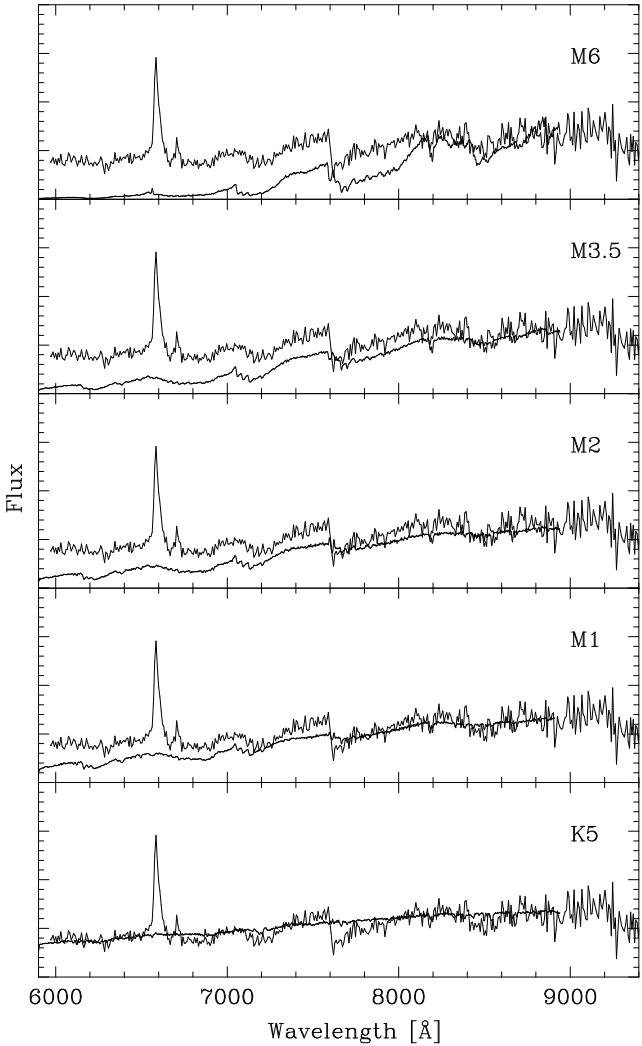}}
\end{picture}
\end{center}
\caption{The eclipse spectrum of MN Hya (which contains
emission lines probably due to some pre-eclipse flux being recorded)
overlaid with various spectral types which were kindly supplied by Dr 
Robert C Smith.}
\label{eclipse} 
\end{figure}

We obtained a red spectrum of MN Hya which was centred on the eclipse
(figure \ref{eclipse}). Although emission lines are visible, this may
be due to some pre or post eclipse flux being recorded. Absorption
features due to the secondary star are visible at 7150 \AA\hspace{1mm}
and 8450 \AA\hspace{1mm}. However, the low resolution of the spectrum
did not permit a detailed analysis of these features. No secondary
features are visible outwith the eclipse.

To determine the spectral type of the secondary star we compared its
spectrum taken during the eclipse with late type spectra kindly
supplied by Dr Robert C Smith. These are overlaid in figure
\ref{eclipse}. The depth of the TiO band at 8450 \AA\hspace{1mm}
indicates that it is later than M1, while the lack of a prominent Na I
feature at 8183--8194\AA\hspace{1mm} suggests that it is earlier than
$\sim$M6.  Bearing in mind that some flux has been recorded which is
not due to the secondary (since we observe emission lines), the
overall spectral shape is closest to M3.5.

Using the empirical mass-radius relationship of Patterson (1984), the
mass of the secondary is expected to be 0.31 $M_{\odot}$ for an
orbital period of $P_{orb}$=3.39 hrs. Using the tables of spectral
type against stellar mass in Zombeck (1990), we find a spectral type
of $\sim$M3. Within the uncertainties involved this is consistent with
the spectral type determined from the eclipse spectrum.

Using the photometry of Sekiguchi, Nakada \& Bassett (1994), where
they find $m_{Ic}$=18.5 at mid-eclipse, and the absolute magnitude of
an M3 V star ($M_{I}$=8.71: Bessell 1991), we obtain a distance of 910
pc if zero absorption is assumed. However, the X-ray data obtained
using {\sl ROSAT} show that the absorption to MN Hya is
$\sim2.5\times10^{20}$ cm$^{-2}$ (Buckley et al 1998b). This implies
$A_{I}$=0.9 and a distance of 610 pc. For a M4 V star ($M_{I}$=9.85:
Bessell 1991) we get distance of 360 and 540 pc assuming $A_{I}$=0.9
and $A_{I}$=0.0 respectively. We estimate that the distance to MN Hya
is between 300--700pc.

\section{Line intensity variations \label{optdip}}

\subsection{Spectra covering the absorption dip}

\begin{figure}
\begin{center}
\setlength{\unitlength}{1cm}
\begin{picture}(8,10.5)
\put(-1.5,-2.5){\includegraphics{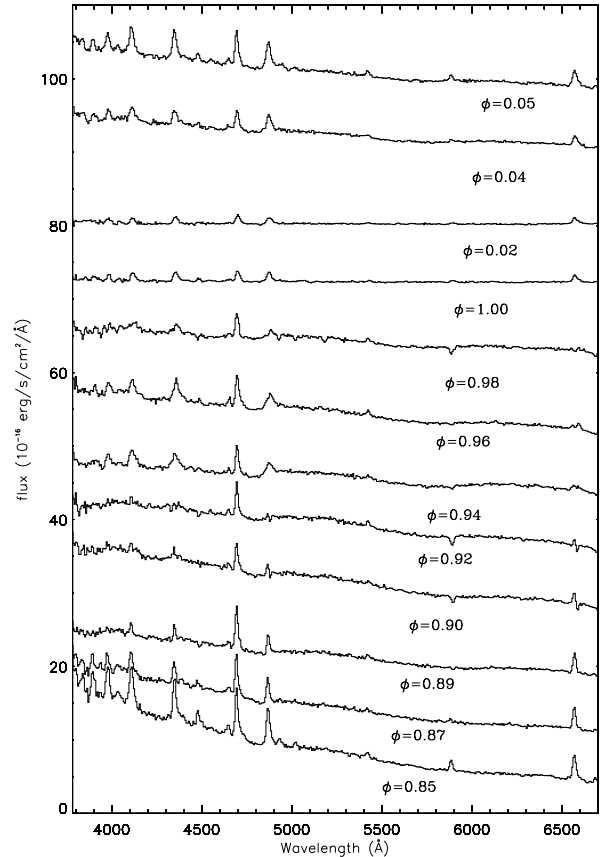}}
\end{picture}
\end{center}
\caption{Spectra starting at $\phi$=0.86 are shown indicating how the
spectrum changes during the course of the absorption dip. Spectra have
been offset by 12 flux units vertically.}
\label{dip} 
\end{figure}

In the previous sections we have concentrated on broad spectral
features. Here we examine individual line variations over the course
of the orbital period.  Perhaps the most interesting feature is the
interval immediately before the eclipse ($\phi\sim$0.90) which
coincides with the most prominent dip seen in soft X-rays (Buckley et
al 1998b). Figure \ref{dip} shows spectra starting at $\phi$=0.85
where the spectrum has strong Balmer and He II line emission and is
typical of a Polar. However, at $\phi$=0.87 the emission lines start
to become less prominent, especially towards bluer wavelengths. At
$\phi$=0.89 only 5 emission lines are present -- HeII $\lambda$ 4100,
H$\delta$, HeI $\lambda$ 4388, HeII $\lambda$ 4686, H$\beta$ and
H$\alpha$. Moving later in phase, at $\phi=0.90$ and 0.92, we find
that HeII $\lambda$ 4686 is still in emission, while H$\alpha$ and
H$\beta$ appear in absorption {\sl and} emission.  At this phase
the emission and absorbed components of the H$\alpha$ line have red
shifted radial velocities of $\sim$270 km s$^{-1}$ and $\sim$870 km
s$^{-1}$ respectively.

Later in phase, at $\phi$=0.94 and 0.96, very broad emission lines are
seen shortwards of 5000 \AA, although HeII $\lambda$ 4686 is as narrow
as before.  In the spectrum taken immediately before eclipse, HeI
$\lambda$ 5876 is back in absorption and all other lines apart from
HeII $\lambda$ 4686 are very weak. By $\phi\sim$0.05, the narrow
component of the lines has returned. In \S 6 we discuss the
interpretation of these dip spectra while in the next section we
examine the line profile variations over the whole of the orbital
period in more detail.

\subsection{Fitting the line profiles}

\begin{figure*}
\begin{center}
\setlength{\unitlength}{1cm}
\begin{picture}(18,10.5)
\put(0,-28.2){\includegraphics{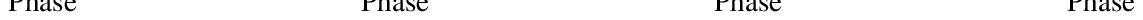}}
\end{picture}
\end{center}
\caption{The line fluxes (10$^{-16}$ erg s$^{-1}$ cm$^{-2}$), radial
velocities (km s$^{-1}$) and EW variations for the H$\alpha$,
H$\beta$, HeI (5876) and HeII (4686) lines over the binary orbital
period. For the H$\alpha$ and H$\beta$ lines we were able to
distinguish a narrow and broad component. We show their combined flux
and EW while we show the radial velocity of the broad component as
a diamond and the narrow component is shown as a plus in the lower panel.}
\label{rvew} 
\end{figure*}

We fitted the emission lines using a single Gaussian profile for the
Helium lines and two Gaussian profiles in the case of the Hydrogen
lines (there were some spectra where an additional Gaussian improved
the fit to the He lines, but not in any consistent manner). The
line-flux was determined by simply summing the flux under the
Gaussian(s). The line-flux variations of H$\alpha$, H$\beta$, HeI
(5876) and HeII (4686) are shown in the top panel of figure
\ref{rvew}.  (We only show the results from the night of Feb 8/9 since
on the previous night the sampling was lower). The radial velocities
and EW for the narrow and broad components of H$\alpha$, H$\beta$,
HeI(5876) and HeII(4686) are shown in figure \ref{rvew}.

The narrow and broad components were fitted with a sinusoidal function
of the form:

\( V(\phi)= \gamma + K \sin(2\pi(\phi-\phi_{o})) \)

\noindent where $V(\phi$) is the radial velocity at phase $\phi$,
$\gamma$ is the component of the accretion streams velocity
perpendicular to the orbital plane projected along the line of sight,
$K$ is the velocity amplitude and $\phi_{o}$ is the phase at which the
observed radial velocity equals the $\gamma$ velocity when crossing
from blue to red. To estimate the errors, the velocity error was
adjusted to normalise the reduced \rchi to 1.0 since it was not
possible to derive meaningful errors to the Gaussian fits. Table
\ref{rvparam} shows the best fit parameters and the 90 per cent
confidence interval while figure \ref{rvew} shows the best fits along
with the data. We now discuss the implications these results have on
our view of the accretion stream.

\begin{table}
\begin{tabular}{lllc}
\hline
Component         & $\gamma$ (km s$^{-1})$& $K$ (km s$^{-1})$ & $\phi_{o}$\\
\hline
H$\alpha$ (narrow)& 102$\pm$29 & 134$\pm$43 & --0.24$\pm$0.04 \\
H$\alpha$ (broad) & 91$\pm$143  & 732$\pm$176 & --0.36$\pm$0.03 \\
H$\beta$ (narrow) & 66$\pm$35  & 259$\pm$52 & --0.28$\pm$0.03 \\
H$\beta$ (broad)  & 102$\pm$75 & 831$\pm$115 & --0.34$\pm$0.02 \\
\hline
\end{tabular}
\caption{The radial velocity parameters for the H$\alpha$ and H$\beta$
lines fitted using two Gaussian profiles. $K$ is the velocity
amplitude and $\phi_{o}$ is the phase at which the observed radial
velocity equals the $\gamma$ velocity when crossing from blue to
red. The errors are quoted at the 90 per cent confidence level.}
\label{rvparam}
\end{table}

\section{The accretion stream}

We have been able to distinguish two components in the Hydrogen line
profiles: a narrow and a broad component. The broad component is most
probably due to emission originating from the accretion stream close
to the white dwarf because of its high maximum radial velocity.  Since
the stream initially has the rotational velocity of the secondary
star, the broad component will be seen with maximum red shift before
$\phi$=0.0 -- in this case at $\phi\sim$0.91. 

The narrow component has a much lower radial velocity than the broad
component: if the narrow component originated from the heated face of
the secondary star we would expect to see it with maximum red shift at
$\phi\sim$0.25. The fact that we see it with maximum red shift at
$\phi\sim$0.98 implies that the narrow component which we can
distinguish is not a good marker of the secondary star. In
non-eclipsing systems, where there is no extra information on the
orbital phase of the secondary, it has often been assumed that the
narrow component originates from the irradiated face of the
secondary. In the case of MN Hya we know that the maximum red-shift of
the narrow component occurs close to the eclipse, so cannot be emitted
by the secondary. Instead the accretion stream close to the secondary
is the most likely candidate and this view is supported by the fact
that the narrow component is eclipsed and with a duration longer than
the white dwarf eclipse. We note that this may well be true for
most/all systems, and that low-resolution spectroscopy is not a
suitable method for detecting the motion of the secondary (and thus
measuring the mass ratio of the system). Schwope, Mantel \& Horne
(1997), for instance, show that with higher resolution spectroscopy, it
is possible to separate the narrow components emitted by the stream
and the secondary.

There are a number of systems which show dips in soft X-rays which
have been attributed to the accretion stream obscuring the hot post
shock region above the white dwarf (eg Watson et al 1989, Buckley et
al 1998b). However, there is only 1 other Polar in which optical lines
appear in absorption during a stream dip (EF Eri: Verbunt et al 1980,
Allen, Ward \& Wright 1981). In the case of EF Eri, lines redder than
5500 \AA\hspace{1mm} appeared in absorption.  In the case of MN Hya we
find that at $\phi\sim$0.90 the H$\alpha$ and H$\beta$ lines are seen
in both emission and absorption, both of which are red-shifted. The
emission and absorption components of the H$\alpha$ line have radial
velocities of $\sim$270 km s$^{-1}$ and $\sim$870 km s$^{-1}$
respectively. Since the stream is curved and we do not know where the
emission and absorption components originate we cannot determine the
true velocities of these components. It is, however, likely that the
absorption component originates closer to the white dwarf than the
emission component. On the other hand, the fact there are two
accreting poles (cf next section) and hence two accretion streams once
the stream attaches onto the magnetic field lines of the white dwarf,
may complicate this interpretation.

\section{Archival Polarisation data}
\label{daveb}

In \S 3 we presented phase resolved optical spectra of MN Hya which
show features which we interpret as cyclotron humps and estimate a
magnetic field strength of the white dwarf to be $B$=42MG from their
spacing.  This is higher than the $B\sim$20MG estimated by Buckley et
al (1998a) from an integrated non-flux calibrated spectrum. Buckley et
al then modeled their white light optical polarimetry using $B$=20
MG. To test if their same polarimetry data can be adequately modeled
using a field strength of $B=$42 MG and determine the location of the
accretion regions on the surface of the white dwarf we measured the
data points from Fig. 4 of Buckley et al (1998a) and fitted their data
using the optimisation method of Potter, Hakala \& Cropper (1998). Up
until now most polarimetry data, including that of Buckley et al, have
been modeled by constructing accretion regions on the surface of the
white dwarf by hand and then a good fit to the data was achieved by
trial and error. The method of Potter, Hakala \& Cropper finds the
best fit in an objective manner. In our new fit we fixed the magnetic
field strength at $B$=42 MG and the inclination at 75$^{\circ}$
(derived from the eclipse duration, Buckley et al 1998a).

\begin{figure}
\begin{center}
\setlength{\unitlength}{1cm}
\begin{picture}(8,11)
\put(-1.5,-3.){\includegraphics{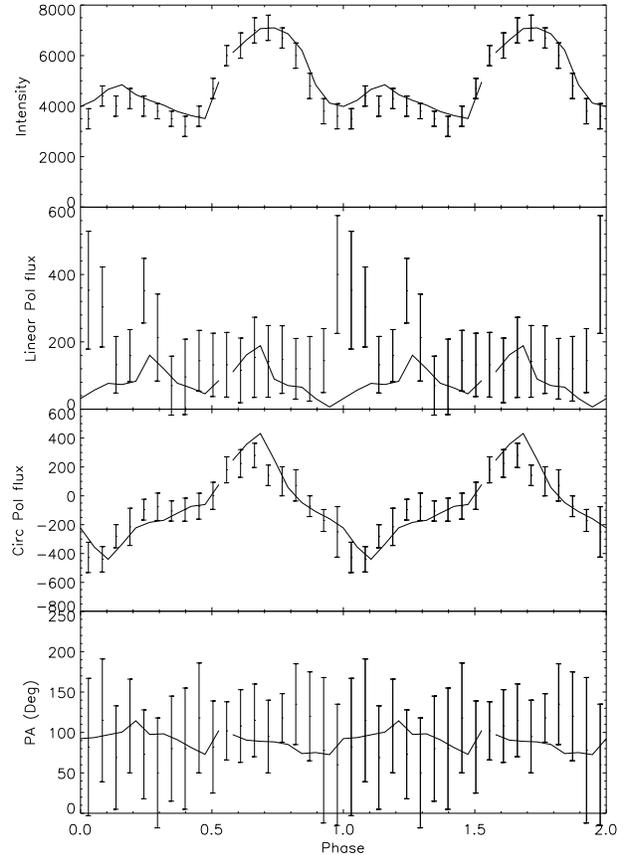}}
\end{picture}
\end{center}
\caption{The white light polarimetry data presented in Buckley et al (1998a) -
from top to bottom: intensity, linearly polarised flux, circularly
polarised flux and the position angle. The best fits using the
optimisation method described by Potter, Hakala \& Cropper 1998 are
overlaid - the inclination was fixed at 75$^{\circ}$, the magnetic
field was fixed at 42 MG.}
\label{steve} 
\end{figure}

The fits of Buckley et al (1998a) implied that two accretion regions
are visible (the main region being partially visible near the lower
magnetic pole) and that they are greatly extended in magnetic
longitude (the secondary region is 300$^{\circ}$ in length and is near
the upper pole). Both accretion regions were offset from their
respective magnetic poles by $\sim20^{\circ}$.  Our fits to the same
polarimetric data are shown in figure \ref{steve}. The fits to the
intensity data (white light) and the circular polarimetry are good and
better than the fit obtained by Buckley et al (1998a). The level of
linear polarisation is rather low and the peak in linear polarisation
around the eclipse phase is almost certainly due to instrumental or
sky background effects in the original data. The best fit to the data
was obtained with the magnetic axis having a tilt of
$\theta_{d}=30^{\circ}$ with respect to the spin axis of the white
dwarf. Further, our fit implies accretion regions which are less
extended than found by Buckley et al: the main region is extended by
less than 60$^{\circ}$ in magnetic longitude and is offset from the
magnetic pole by 15$^{\circ}$ and is visible on the limb of the white
dwarf between $\phi\sim0.45-0.75$. The secondary region (which is much
less bright than the primary region) is roughly circular in shape and
is visible between $\phi\sim0.65-0.25$ and comes almost face on at
$\phi$=0.9.

The location of the primary accretion region is consistent with the
detection of cyclotron humps in the optical spectra between
$\phi\sim0.35-0.65$. The fact that no cyclotron features are seen when
the secondary accretion region is visible may simply be due to the
features being too faint to detect or that the magnetic field at the
secondary region is lower than the primary accretion region. Buckley
et al (1998b) suggested that the latter may account for the higher
soft to hard X-ray ratio over $\phi$=0.03--0.45 compared to
$\phi$=0.45--0.87.

We conclude that our estimate of $B=42$MG is consistent with the
optical polarimetry and that at $\phi\sim$0.9 it is the secondary
accretion region which is obscured by the accretion stream.

\section{Acknowledgments}

This paper is based on observations obtained at the European Southern
Observatory, La Silla, Chile using the ESO 3.6m telescope and we would
like to thank ESO for their support during our observing period. PJW
is supported by a PPARC post-doctoral fellowship. We are grateful to
Dr Stephen Potter for modelling the photo-polarimetric data of Buckley
et al and Dr Robert C. Smith for supplying us with spectra of
late-type stars.  Further, thanks go to Dr David Buckley for suppling
the eclipse ephemeris and pre-prints before publication.

\end{document}